\documentclass[smallextended]{svjour3}

\smartqed

\usepackage{bm}
\usepackage{amsfonts}
\usepackage{graphicx}

\journalname{General Relativity and Gravitation}

\begin{document}

\title{Dark Energy and Extending the Geodesic Equations of Motion:}
\subtitle{Its Construction and Experimental Constraints}  

\titlerunning{Extending the GEOM}

\author{A.~D.~Speliotopoulos}

\institute{
Department of Physics
University of California
Berkley, CA 94720, and 
Department of Mathematics,
Golden Gate University,
San Francisco, CA 94105,
ads@berkeley.edu
}

\date{November 16, 2009}

\maketitle

\begin{abstract}

With the discovery of Dark Energy, $\Lambda_{DE}$, there is now a universal
length scale, $\ell_{DE}=c/(\Lambda_{DE} G)^{1/2}$, associated with
the universe that allows for an extension of the geodesic equations of
motion. In this paper, we will study a specific class of such
extensions, and show that contrary to expectations, they are not
automatically ruled out by either theoretical considerations or
experimental constraints. In particular, we show that while these
extensions affect the motion of massive particles, the motion of
massless particles are not changed; such phenomena as gravitational
lensing remain unchanged. We also show that these extensions do not
violate the equivalence principal, and that because
$\ell_{DE}=14010^{800}_{820}$ Mpc, a specific choice of this
extension can be made so that effects of this extension are not be
measurable either from terrestrial experiments, or through observations
of the motion of solar system bodies. A lower bound
for the only parameter used in this extension is set.  

\PACS{95.36.+x\and 04.20.Cv\and 04.25.-g}
 
\end{abstract}


\section{Introduction}
\label{intro}

The recent discovery of Dark Energy (see \cite{Ries1998}, \cite{Perl1999}
and references therein) has broadened our knowledge of the universe, and
has demonstrated once again that it can hold surprises for
us. This discovery has, most assuredly, also brought into 
sharp relief the degree of our understanding of the universe. In this
paper, we will study one specific implication of this discovery:  With
the discovery of Dark Energy, $\Lambda_{DE}$, there is now a
\textit{universal} length scale, $\ell_{DE} =
c/(\Lambda_{DE}G)^{1/2}$, \footnote{Our $\ell_{DE}$
  differs from the length scale, $r_\Lambda$, defined in \cite{Bous}
  by a factor of $\sqrt{3}$.}, \footnote{It is also possible to construct
  the length scale $(\hbar c/\Lambda_{DE})^{1/4}\approx 85$ $\mu$
  m. Experiments have shown that this scale does not affect the
  Newtonian potential \cite{Adel2007}.} associated with the universe
that allows for extensions of the geodesic equations of motion (GEOM). We
find that contrary to expectations, these extensions are not
automatically ruled out by theoretical considerations, nor are they
ruled out by experimental constraints either through terrestrial
experiments or through solar system tests of general
relativity. Indeed, we show in this paper that one specific extension
of the GEOM is a viable alternative to the GEOM, and we obtain a lower
bound for the only free parameter used in its construction, a
power-law exponent, $\alpha_\Lambda$.   

There are good theoretical and physical reasons for studying the
range of extensions of the geodesic equations of motion that are
allowed. Arguments for the use of the geodesic equations of motion to
describe the motion of massive test particles in curved spacetime are
based on various statements of the equivalence principle, and the
principle of general covariance (see chapter 4 of \cite{Wald}), along
with arguments in favor of simplicity and aesthetics. Importantly, these
arguments are made in addition to those made in favor of Einstein's
field equation. Namely, these is no unique way of deriving the
geodesic equations of motion from the field equations.  Indeed, in
1938 Einstein, Infield and Hoffman attempted to show that as a
consequence of the field equations, massive test particles will travel
along geodesics in the spacetime \cite{Einstein}. These attempts have
continued to the present day \cite{Gero1972}, \cite{Gero2003}.

Extensions and modifications of the GEOM have been made before, of
course. On the level of Newtonian dynamics, Modified Newtonian
Dynamics (MOND) \cite{MOND} has been proposed as an alternate
explanation of the galactic rotation curves. On the relativistic
level, there has been recent efforts \cite{Lamm} to develop a general
framework to study modifications to the GEOM in the weak field,
linearized gravity limits. The major impetus for this work has been to
describe a series of dynamical anomalies\textemdash the
Pioneer anomaly (see \cite{Ande1998} and \cite{Ande2002}), the flyby
anomaly \cite{Ande2001}, and the lengthening of the Astronomical Unit
\cite{Kras}\textemdash that have been observed at the Solar system scale.  

The focus of this paper is to establish the underlying theoretical
framework that can be used to describe structures and dynamics at the
galactic length scale and above. In a future paper \cite{ADS}, this
framework will be applied to an analysis of the galactic rotation
curves, and the impact that this extension has on
phenomena at cosmological length scales will be studied.  As such, we
focus here on the Dark Energy energy length scale, and on how the
existence of this scale allows for extensions of the GEOM. Indeed, we
find that with this length scale, $\ell_{DE}$, extensions of the GEOM
are not difficult to construct. The quotient $c^2R/\Lambda_{DE}G$ is
dimensionless, and functionals of this quotient can easily be used to
extend the GEOM. What is more relevant is whether or not  
the resultant equations of motion will be a physically viable
alternative to the GEOM. As such we will be guided in our extension
of the GEOM by the four conditions listed below. They are deliberately
chosen to be conservative in scope, and thus stringent in their
application. Somewhat surprisingly, we will show that there is at
least one extension of the GEOM that satisfies all four.  

First, we require that the extension preserve the equivalence
principle, which is one of the underlying principles on which general
relativity is founded. In the following sections of the paper, we will
explicitly see that this preservation is assured by the fact that
$\ell_{DE}$ is the same for all test particles. This universal nature
of $\ell_{DE}$ is crucial. While other length scales\textemdash say,
the proton mass\textemdash could be used for the extension, the
resultant equations of motion would depend on this mass. They could
not then be applied to the motion of protons without explicitly
violating the uniqueness of free fall condition. 

Second, we require that the extension not change the
equations of motion for massless test particles; such particles must
still follow the GEOM. All astronomical observations\textemdash of
which gravitational lensing is playing an increasingly important
role\textemdash are based on the motion of photons of various
wavelengths. Modifications to the equations of motion for photons will 
require a reinterpretation of these observations, a daunting step not
to be taken without good reason. We will show that by considering a
class of extensions that is based on conformally scaling the rest
mass of the test particle, we arrive at extended GEOMs that, on the
one hand, will not change the motion of massless test particles, but 
will, on the other, change the motion of massive ones. Being
conformal, the motion of photons will not be affected by this class
of extensions, and they will still travel along null geodesics; phenomena
such as gravitational lensing will remain unchanged. While in form this
class of extensions resembles a scalar field theory that is
non-minimally and  nonlinearly coupled to the scalar curvature,
$R$, such theories are constructed at the quantum level. Our extension of
the GEOM is done at the \textit{classical}, $\hbar = 
0$, level, with the scale of the coupling set by $\ell_{DE}$.

The third condition involves the attempts \cite{Einstein},
\cite{Gero1972}, \cite{Gero2003} at proving that the GEOM are the
unique consequence of the Einstein field equations (see also page 72
of \cite{Wald}). These proofs would seem to rule out any physically
relevant extension of the GEOM, and by necessity, our extension of
the GEOM cannot be precluded by such proofs. That the extension is possible is
because these proofs focus on the motion of test particles in regions
where the Einstein tensor, $G_{\mu\nu}$, vanishes. We will see that in
these regions the extended GEOM reduce to the GEOM, and thus do not
violate these proofs. Indeed, we will explicitly construct the
energy-momentum tensor for an inviscid fluid of massive particles
propagating under the extended GEOM.   

The fourth condition is the most stringent of the four. With the
exception of the as-yet unexplained anomalies described above, we
require that the extension of the GEOM not produce effects that are
measurable either in terrestrial experiments, or through the 
motion of bodies in the solar system that have traditionally been used
to test general relativity. While stringent, we will nevertheless show
explicitly that a choice of extension can be made which
satisfies it. Physically, this choice is possible because at
$(7.21^{+0.83}_{-0.84}) \hbox{ x } 10^{-30}$ $g/cm^3$ \cite{WMAP},
$\Lambda_{DE}$ is orders of magnitude smaller than the density of
matter, $\rho_{\hbox{\scriptsize{limit}}}$, either currently
achievable in terrestrial experiments (where densities exceed
$10^{-18}$ g/cm$^3$), or present in the solar system  (where the
density of matter in Mercury's orbit exceeds $10^{-23}$
g/cm$^3$). Correspondingly, at $14010^{800}_{820}$ Mpc $\ell_{DE}$ is
more than three times larger than the observed size of the universe,
and is orders of magnitude larger than the solar system. Nevertheless,
we find that even though the disparity between the magnitude of
$\Lambda_{DE}$ and $\rho_{\hbox{\scriptsize{limit}}}$\textemdash or,
equivalently, between $\ell_{DE}$ and the size of the solar
system\textemdash is large, a nonlinear function of
$c^2R/\Lambda_{DE}G$ is needed in 
constructing the extension for its effects not to have already
been seen in terrestrial experiments. The simplest of these extensions
has only one free parameter, $\alpha_\Lambda$, a power-law exponent
that determines the behavior of the function at densities both much
larger than $\Lambda_{DE}$, and much smaller than it. Lower bounds for
$\alpha_\Lambda$ are set by requiring that the extension does not
produce observable effects in current terrestrial experiments. 

While it may be possible to apply the analysis in this paper to the
explanation of Solar system anomalies such as the Pioneer anomaly, the
focus of this research is on phenomenon at the galactic scale or
longer. It is for this reason that we require our extension to be
constrained only by experiments and observations that are currently
well-understood, and for which the underlying physics is well-known.
We leave to future work the question of whether or not our analysis
can be applied to explaining the Pioneer and other Solar-system-scale
anomalies.   

\section{Extending the Geodesic Lagrangian}
\label{sec:1}

We begin our extension of the GEOM with Einstein's field equation in
the presence of a cosmological constant 
\begin{equation}
  R_{\mu\nu} - \frac{1}{2}g_{\mu\nu}R +
  \frac{\Lambda_{DE}G}{c^2} g_{\mu\nu}= - \frac{8\pi G}{c^4} T_{\mu\nu},
\label{EinsteinEquation}
\end{equation}
where $T_{\mu\nu}$ is the energy-momentum tensor for matter,
$R_{\mu\nu}$ is the Ricci tensor, Greek indices run from $0$ to $3$,
and the signature of $g_{\mu\nu}$ is $(1,-1,-1,-1)$. While
there is currently no consensus as to the nature of Dark
Energy (proposals have been made that identify it with the
cosmological constant $\Lambda_{DE}$ \cite{Bous}, with quintessence
\cite{Peebles1988}, \cite{Stei1999}, \cite{Stei2000}, or even as a
consequence of loop quantum cosmology \cite{Cahi}),
modifications to Einstein's equations to include the cosmological
constant are well known and are minimal. We will thus identify Dark
Energy with the cosmological constant in this paper, and require only 
that $\Lambda_{DE}$ changes so slowly that it can be considered a
constant in our analysis. 

Requiring that Eq.~$(\ref{EinsteinEquation})$ still holds under the
extension of the GEOM is a \textit{choice}, one which, we will see 
below, is the simplest. Although it may seem surprising that we can
still make this choice even though we will be changing the GEOM,
extensions of the GEOM need not change the relation between
$R_{\mu\nu}$ and $T_{\mu\nu}$ given in
Eq.~$(\ref{EinsteinEquation}$). They can rather change the precise  
form that $T_{\mu\nu}$ takes for matter. To see this, consider the
following.  

The total action, $S$, for a system consisting of gravity, radiation,
and matter can be written as a sum of three parts: $S=
S_{\hbox{\scriptsize{grav}}} + S_{\hbox{\scriptsize{radiation}}}
+S_{\hbox{\scriptsize{matter}}}$. Here, 
$S_{\hbox{\scriptsize{grav}}}$ is the action for gravity, 
$S_{\hbox{\scriptsize{radiation}}}$ is the action for radiation, and
$S_{\hbox{\scriptsize{matter}}}$ is the action for matter. We will
show below that the class of extended GEOM we consider here will not
change the equations of motion for radiation so that
$S_{\hbox{\scriptsize{radiation}}}$ will not be changed. The
extension will certainly change $S_{\hbox{\scriptsize{matter}}}$,
however, and in the next section we will explicitly construct the
energy-momentum tensor for an inviscid fluid whose constituents follow
an extended GEOM. For the present argument, we only need to note that whatever
the form taken for $S_{\hbox{\scriptsize{matter}}}$, we are still free
to choose $S_{\hbox{\scriptsize{grav}}}$ to be the Hilbert action; 
Eq.~$(\ref{EinsteinEquation})$ then follows after taking the  
functional derivative of $S$ with respect to the metric. 

This ability to change $S_{\hbox{\scriptsize{matter}}}$ while leaving
$S_{\hbox{\scriptsize{grav}}}$, and thus
Eq.~$(\ref{EinsteinEquation})$, unchanged was explicitly
exploited in the construction of minimally coupled scalar fields,
$\phi_R$. There, the mass term of the scalar field, $m^2\phi_R^2$, is
replaced by $Rm^2\phi_R^2$ in the action for matter, and yet
$S_{\hbox{\scriptsize{grav}}}$ is still taken to be the Hilbert action (see
Sec. \textbf{II.D}). Einstein's field equations,
Eq.~$(\ref{EinsteinEquation})$, still hold; the only change is the
form that $T_{\mu\nu}$ takes.   

Both the geodesic Lagrangian
\begin{equation}
\mathcal{L}_0 \equiv mc \left(g_{\mu\nu}\frac{d x^\mu}{dt}\frac{d
  x^\nu}{dt}\right)^{1/2},
\label{geoL}
\end{equation}
and the GEOM
\begin{equation}
v^\nu\nabla_\nu v^\mu \equiv \frac{D v^\mu}{\partial t} = 0,
\label{geoEOM}
\end{equation}
(where $v^\mu = \dot{x}^\mu$ is the four-velocity of the test
particle), have natural geometric meaning. The first is a 
proper time interval, while the second is the equation for parallel
transport, which determines the shortest time-like path connecting
two points. But, aside from their inherent geometric meaning, there is
also a good physical reason to take Eq.~$(\ref{geoEOM})$ as the
equations of motion for a test particle. In the absence of Dark Energy,
Eq.~$(\ref{geoEOM})$ is the most general form that a second-order
evolution equation for a test particle can take which still obeys the
equivalence principle.  

Any extension of $\mathcal{L}_0$ would require a dimensionless, 
scalar function of some fundamental property of the spacetime 
folded in with some physical property of the universe. In our
homogeneous and isotropic universe, there are few 
opportunities to do this. A fundamental vector certainly does not
exist in the spacetime, and while there is a scalar (the scalar
curvature, $R$) and three tensors ($g_{\mu\nu}$, the Riemann tensor,
$R_{\mu\nu,\alpha}^{\>\>\>\,\quad\beta}$, and the Ricci tensor, $R_{\mu\nu}$),
$R_{\mu\nu,\alpha}^{\>\>\>\,\quad\beta}$  has units of inverse length
squared. While it is possible to construct a dimensionless scalar
$m^2 G^2 R/c^4$ for the test particle, augmenting $\mathcal{L}_0$ using a
function of this scalar would introduce additional forces that will
depend on the mass of the test particle, and thus violate the
uniqueness of free fall principle. It is also possible to construct
the scalar $g_{\mu\nu}v^\mu v^\nu/c^2$, but because of the mass-shell
condition, $v_\mu v^\mu=c^2$, any such extension of $\mathcal{L}_0$
will not change the GEOM. Scalars may also be constructed from
$R_{\mu\nu}$ and powers of $R_{\mu\nu,\alpha}^{\>\>\>\,\quad\beta}$ by
contracting them with the appropriate number of $v^\mu/c$'s, but these
scalars will once again have dimension of inverse length raised to
some power, and, as with the Ricci scalar, once again a rest mass $m$
is needed to construct the dimensionless quantity. 

The situation changes dramatically in the presence of Dark Energy. With a
universal length scale, $\ell_{DE}$, it is now possible to construct
from the Riemann tensor and its contractions dimensionless scalars of
the form, 
\begin{equation}
\frac{c^2R}{\Lambda_{DE}G},\>\>\> \frac{R_{\mu\nu}v^\mu
  v^\nu}{\Lambda_{DE}G}, \> \>\>\frac{c^2v^\mu
  v^\nu}{(\Lambda_{DE}G)^2} R_{\mu{\alpha}, \beta\gamma}
  R_\nu^{\>\>\>{\alpha}, \beta\gamma}, \>\>\>
  \frac{v^\mu v^\nu
  v_\gamma v_\delta}{(\Lambda_{DE}G)^2}R_{\mu{\alpha}, \nu\beta}
  R^{\gamma {\alpha}, \delta \beta}, \>\>\> \dots .
\label{possibles}
\end{equation} 
Although extensions to $\mathcal{L}_0$ can be constructed with any of
these terms, we look at extensions with the form: 
\begin{equation}
\mathcal{L}_{\hbox{\scriptsize{Ext}}} \equiv
mc\Big[1+\mathfrak{D}\left(c^2R/\Lambda_{DE}G\right)\Big]^{\frac{1}{2}}
\left(g_{\mu\nu}\frac{d x^\mu}{dt}\frac{d x^\nu}{dt}\right)^{\frac{1}{2}} \equiv
\mathfrak{R}[c^2R/\Lambda_{DE}G] \mathcal{L}_0,
\label{extendL}
\end{equation} 
with the implicit condition that $v^2=c^2$ for massive test
particles. We make this choice for the following reasons. 

First, having $\mathfrak{R}$ be a function only of
$c^2R/\Lambda_{DE}G$ is the simplest extension that can be chosen;
other choices will induce velocity-dependent effects in the extended
GEOM. Second, we will find below that \textit{any} extension of
$\mathcal{L}_0$ of the form Eq.~$(\ref{extendL})$ will
\textit{not} change the equations of motion for massless test particles;
they can still be reduced to the GEOM. Extensions of the form
Eq.~$(\ref{extendL})$ will only affect the motion of massive
test particles. Third, we require that the extension of the GEOM 
not produce effects that should have already been seen in terrestrial
experiments; these experiments are done in the nonrelativistic and
weak gravity limits. Constraints in the choice of $\mathfrak{R}$ are
thus found in these limits, where it is clear from
Eq.~$(\ref{possibles})$ that the second term reduces to the
first, while the other terms are higher order in the curvature. We are
thus left with $c^2R/\Lambda_{DE}G$ with which to construct an
extension of the GEOM.  

As $\ell_{DE}=14010^{800}_{820}$ Mpc, the question remains whether it
is possible to use another, shorter length scale in its place to
extended the GEOM; this extension could then be used to describe
deviations from geodesic motion on shorter length scales. One 
such application would be in explaining Solar system scale anomalies
such as the Pioneer and flyby anomalies.  At this scale, a natural
length scale would be $M_\odot G/c^2=1480$ m, where $M_\odot$ is the
mass of the Sun, and the resultant extension of the GEOM may be
applicable to the description of motion on the Solar system scale (see
\cite{Lamm}).  Its application to the description of motion
at the galactic scale or longer is more  problematic, however, and it
is precisely on these length scales that we are concerned with here.  On
the galactic scale, stars can be treated as test particles, and as
$M_\odot G/c^2$ depends explicitly on the mass of the Sun, the use of
this length scale in extending the GEOM would mean that the motion of
stars in galaxies would depend on the mass of the Sun. This would not
be physically reasonable, and would also violate the uniqueness of
free fall condition.   

If $\mathfrak{D}(x)$ is the constant function, then
$\mathcal{L}_{\hbox{\scriptsize{Ext}}}$ differs from $\mathcal{L}_0$
by an overall constant that can be absorbed through a
reparametrization of time. Only non-constant $\mathfrak{D}(x)$ are
relevant. It is how \textit{fast} $\mathfrak{D}(x)$ changes that will
determine its effect on the equations of motion, and not its overall
scale. Indeed, in extending $\mathcal{L}_0$ we have in effect
performed a conformal scaling of $\mathcal{L}_0$ 
by replacing the constant rest mass $m$ of the test particle with a
curvature-dependent rest mass $m\mathfrak{R}\left[c^2R/
  \Lambda_{DE}G\right]$. All \textit{dynamical} effects of this
extension can therefore be interpreted as the rest energy gained or
lost by the test particle due to the local curvature of the
spacetime. The scale of these effects is of the order of $mc^2/L$,
where $L$ is some relevant length scale of the dynamics. The
additional forces from $\mathcal{L}_{\hbox{\scriptsize{Ext}}}$ are
thus potentially very \textit{large}. For these effects \textit{not} to
have already been seen, $\mathfrak{D}(c^2R/\Lambda_{DE}G)$ must 
change very slowly at current limits to experimental measurements.

As mentioned above, using Einstein's field equations,
Eq.~$(\ref{EinsteinEquation})$, was a choice. In particular, notice that
because $\mathcal{L}_{\hbox{\scriptsize{Ext}}}$ is the result of
conformally scaling the rest mass by $\mathfrak{R}$, we may choose to
instead reduce $\mathcal{L}_{\hbox{\scriptsize{Ext}}}$ to
$\mathcal{L}_0$ through the conformal transformation of the metric
$\tilde{g}_{\mu\nu} = \mathfrak{R}^{1/2}g_{\mu\nu}$. However, 
doing so will result in a Ricci tensor $\tilde{R}_{\mu\nu}$ that is
nonlinearly related to 
$R_{\mu\nu}$: 
\begin{eqnarray}
\tilde{R}_{\mu\nu} = &&R_{\mu\nu}
-\frac{1}{2}\nabla_\mu\nabla_\nu\log{\mathfrak{R}}
  -\frac{1}{4} g_{\mu\nu}\nabla_\alpha\nabla^\alpha\log{\mathfrak{R}}
\nonumber 
\\
&&+\frac{1}{8}\nabla_\mu\log{\mathfrak{R}}\nabla_\nu\log{\mathfrak{R}}-\frac{1}{8}g_{\mu\nu}\nabla_\alpha\log{\mathfrak{R}}\nabla^\alpha\log{\mathfrak{R}}.
\end{eqnarray}
where $\nabla_\mu$ is the covariant derivative for
$g_{\mu\nu}$. As Eq.~$(\ref{EinsteinEquation})$ holds for
$R_{\mu\nu}$, it cannot hold for $\tilde{R}_{\mu\nu}$. It will instead
be replaced by a nonlinear relation between $\tilde{R}_{\mu\nu}$ and
$T_{\mu\nu}$, resulting in a higher-order theory of gravity. Thus,
instead of choosing Eq.~$(1)$ to hold for the Ricci tensor 
and extend the GEOM, we could have chosen a higher order theory of
gravity from the start while preserving the GEOM. We would 
argue, however, that making this second choice would change the
foundations of classical general relativity\textemdash a much more
drastic step\textemdash that would also result in a less tractable
theory. 

\subsection{The Extended GEOM for Massive Test Particles}
\label{sec:2}

For massive test particles, the extended GEOM from
$\mathcal{L}_{\hbox{\scriptsize{Ext}}}$ is
\begin{equation}
\frac{D^2 x^\mu}{\partial t^2} = c^2 \left(g^{\mu\nu} - \frac{v^\mu
  v^\nu}{c^2}\right)
  \nabla_\nu\log\mathfrak{R}[c^2R/\Lambda_{DE}G],
\label{genEOM}
\end{equation}
where we have explicitly used $v^2=c^2$. It has a canonical momentum
with a magnitude of
\begin{equation}
p^2 = p_\mu p^\mu =m^2c^2 \bigg[
1+\mathfrak{D}(c^2R/\Lambda_{DE}G)\bigg],
\label{extendMass}
\end{equation}
and the interpretation of $m\mathfrak{R}[c^2R/\Lambda_{DE} G]$ as an
effective rest mass can be readily seen. What also can be readily seen
is that $\mathcal{L}_{\hbox{\scriptsize{EXT}}}$ and Eq.~$(\ref{genEOM})$
have lost the geometrical meaning that $\mathcal{L}_0$ and
Eq.~$(\ref{geoEOM})$ have. Namely, the worldline of a massive test particle
is not that which minimizes the proper time between two points; it is
instead one that is either attracted to, or repelled from (depending on
the choice of $\mathfrak{R}$), regions where the scalar curvature is
extremized.   

The dynamical implications of the new terms in Eq.~$(\ref{genEOM})$,
along with the conditions under which they are relevant, can most
easily be seen by noting that $R = 4 \Lambda_{DE}G/c^2 +8\pi
TG/c^4$, where $T = T_\mu^\mu$. Then
$\mathfrak{R}[c^2R/\Lambda_{DE}G]=\mathfrak{R}[4+8\pi T/\Lambda_{DE}
  c^2]$, where the `4' comes from the dimensionality of
spacetime. \textit{Thus, in regions of spacetime where either $T_{\mu\nu} =0$
or when $T$ is a constant, the right hand side of
Eq.~$(\ref{genEOM})$ vanishes, and our extended GEOM reduces back to
the GEOM.}

\subsection{Dynamics of Massless Test Particles}
\label{sec:3}

For massless test particles, we note that
$\mathcal{L}_{\hbox{\scriptsize{Ext}}}$ is related to $\mathcal{L}_0$
by a scaling of the rest mass. This scaling can be  formally
interpreted as a conformal transformation of the metric,
$g_{\mu\nu}$.  As null geodesics are preserved under conformal
transformations up to an affine transformation, the extended GEOM is
equivalent to the GEOM, and we still have $v^\nu \nabla_\nu v^\mu =
0$. 

This result can be shown explicitly in the following analysis. Because
massless particles follow null geodesics, for these particles
we consider extensions of the form   
\begin{equation}
\mathcal{L}_{\hbox{\scriptsize{Ext}}}^\gamma=\mathfrak{R}[c^2R/\Lambda_{DE}G]
\mathcal{L}_0^\gamma,
\label{ExtLagranMassless}
\end{equation}
where we have taken
\begin{equation}
\mathcal{L}_0^\gamma = \frac{1}{2}g_{\mu\nu}\frac{dx^\mu}{dt}\frac{dx^\nu}{dt}.
\end{equation}
As usual, the GEOM comes from $\mathcal{L}^\gamma_0$, while the equations of
motion that come from $\mathcal{L}_{\hbox{\scriptsize{Ext}}}^\gamma$ are
\begin{equation}
0=\mathfrak{R}\left(\frac{d\>\>\>}{dt}\frac{\delta\mathcal{L}_0^\gamma}{\delta
  \dot{x}^\mu} -\frac{\delta \mathcal{L}_0^\gamma}{\delta
  x^\mu}\right) +\frac{\delta\mathcal{L}_0^\gamma}{\delta \dot{x}^\mu}
\frac{d\mathfrak{R}}{dt} - \mathcal{L}_0^\gamma\nabla_\mu\mathfrak{R}.
\label{minMassless}
\end{equation}  
Taking now the mass-shell condition, $\mathcal{L}_0^\gamma=0$,
Eq.~$(\ref{minMassless})$ reduces to simply
\begin{equation}
0=\frac{D\>\>\>}{\partial t}
\left(\mathfrak{R}\frac{dx^\mu}{dt}\right). 
\label{masslessEOM}
\end{equation}  
After the reparametrizing $dt \to \mathfrak{R}dt$ \cite{Wald}, we
arrive at the expected result, $v^\nu\nabla_\nu v^\mu=0$. Importantly,
this result means that the usual general relativistic effects associated
with photons\textemdash the gravitational redshift and the deflection
of light\textemdash are not effected by our extension of the GEOM.     

\subsection{Impact on the Equivalence Principles}
\label{sec:4}

The statements \cite{MTW} of the equivalence principal we are
concerned with here are the following: 

\vskip 10pt
\noindent{\textit{Uniqueness of Free Fall:} It is clear from
  Eq.~$(\ref{genEOM})$ that the worldline of a freely falling test particle
  under the extended GEOM does not depend on its composition or
  structure. 

\vskip 10pt
\noindent{\textit{The Weak Equivalence Principle:}  Our extension also
  satisfies the weak equivalence principle to the same level of
  approximation as the GEOM. The weak equivalence principle is
  based on the ability to choose a frame in a neighborhood of
  the worldline of the test particle where $\Gamma^\mu_{\alpha\beta}
  \approx0$; the Minkowski metric, $\eta_{\mu\nu}$, is then a good
  approximation to $g_{\mu\nu}$ in this neighborhood. However, as one
  deviates from this world line corrections to $\eta_{\mu\nu}$ appear,
  and since a specific coordinate system has been chosen, they appear
  as powers of the Riemann tensor (or its contractions), and its
  derivatives (see \cite{MTW} and \cite{Fermi}). This means that the
  \textit{larger} the curvature, the \textit{smaller} the neighborhood
  about the world line where $\eta_{\mu\nu}$ is a good approximation
  of the metric. Consequently, the weak equivalence principle holds up
  to terms first order in the curvature.  As the additional
  terms in Eq.~$(\ref{genEOM})$ are first order in $R$ as well, our
  extension of the GEOM satisfies the weak equivalence principle to
  the same order of approximation as the GEOM does. 
\vskip 10pt

\noindent{\textit{The Strong Equivalence Principle:} Because we only
  change the geodesic Lagrangian, all nongravitational 
forces in our theory will have the same form as their special relativistic
counterparts. Moreover, the extended GEOM reduces to the GEOM in the
$R\to 0$ limit. 

\subsection{Connections with Other Theories}
\label{sec:5}

As unusual as the extended GEOM, Eq.~$(\ref{genEOM})$, may appear
to be, there are connections between this extension and other theories. 

\subsubsection{The Class of Scalar Field Theories in Curved Spacetimes}
\label{sec:6}

The Klein-Gordon equation corresponding to the extended GEOM is 
\begin{equation}
\nabla^2 \phi + \frac{m^2c^2}{\hbar^2}\left[
  1+\mathfrak{D}\left(\frac{c^2R}{\Lambda_{DE} G}\right)
\right]\phi = 0.
\label{extendKG}
\end{equation}
This is the equation of motion for a scalar field $\phi$ that is
non-minimally and nonlinearly coupled to $R$. Scalar field theories of
this class have been studied before, the most notable of which is  
\begin{equation}
\nabla^2\phi_{R} + \left(\frac{m^2c^2}{\hbar^2}+\xi R\right)\phi_R=0.
\label{conformalS}
\end{equation}
When $\xi = 1/6$, the scalar field will be conformally
invariant even though $m\ne 0$ \cite{BD}. 

There are important
differences between these theories and the theory we are considering
here, however. Scalar field theories of the form
Eq.~$\ref{conformalS})$ were proposed at the quantum level and $\hbar$
appears explicitly; we are focused on the classical, $\hbar=0$,
level. Note also that the \textit{scale} of the  coupling in
Eq.~$(\ref{extendKG})$ is set by $\ell_{DE}$. This is a macroscopic
in length scale, and if we expand $R$ about $4\Lambda_{DE}G/c^2$,
Eq.~$(\ref{extendKG})$ reduces to Eq.~$(\ref{conformalS})$ with a 
\begin{equation}
\xi = \mathfrak{D}'(4)\frac{m^2c^2\ell_{DE}^2}{\hbar^2},
\end{equation}
which has a magnitude $\sim
10^{75}$\textemdash indicative of an inherently classical
theory\textemdash if $\phi$ has the mass of a proton. This
value for $\xi$ is \textit{seventy-five} orders of magnitude larger than
the values of $\xi$ usually considered. It also signifies that a 
perturbative solution of Eq.~$(\ref{extendKG})$ would be of limited
use at best, and the non-linearity of the coupling must be explicitly
taken into account. 

\subsubsection{The $f(R)$ Theory}
\label{sec:7}

Proposals for modifying the Hilbert action by considering functions, $f(R)$,
of the Ricci scalar have been made before (see \cite{Nav-2006-1} and
\cite{Noji-2006} for reviews). These theories were first introduced to
explain cosmic acceleration without the need for Dark Energy
using a $1/R$ action \cite{Cap}, \cite{Turner}, and further extensions of
this model have been made \cite{Noji-2003-1}, \cite{Noji-2007} since
then. They are now being studied in their own right, and various
functional forms for $f(R)$ are being considered. Indeed, connection
to Modified Newtonian Dynamics has been made for logarithmic $f(R)$
terms \cite{Noji-2003-2}, \cite{Nav-MOND}, while with other 
choices of $f(R)$ connection with quintessence has been made
\cite{Nav-2006-2} - \cite{Chib} as well. Importantly, issues 
with the introduction of a ``fifth force'', and compatibility with
terrestrial experiments have begun to be addressed through the
Chameleon Effect (see \cite{Khou-1}- \cite{Mota} and an overview in
\cite{Nav-2006-2}), which is used to hide the effects of field with a
small mass that would otherwise be seen.  

It is also important to note that while $f(R)$ theories change the
action for gravity, in our approach we do not; we still take the
action for gravity to be the Hilbert action with the addition of a
cosmological constant. We instead change the \textit{response} of matter to
gravity by extending the equations of motion for test particles, and
thus change the energy-momentum tensor for matter.

\section{The Energy-Momentum Tensor}
\label{sec:8}

Beginning with \cite{Einstein}, there have been a number of
attempts to show that the GEOM are a necessary \textit{consequence} of
the Einstein's field equations, Eq.~$(\ref{EinsteinEquation})$. Modern
attempts at demonstrating such a linkage \cite{Gero1972},
\cite{Gero2003} focus on the energy-momentum tensor, and consider the 
motion of a test particle moving in a region of spacetime where
$T_{\mu\nu}=0$ outside of a ``worldtube'' that surround the test
particle; inside this worldtube, Einstein tensor $G_{\mu\nu} \ne
0$. In fact, this tensor must satisfy the strong energy condition
$G_{\mu\nu}t^\mu {t'}^\nu \le 0$ (for our signature for the metric) there,
where $t^\mu$ and ${t'}^\nu$ are two arbitrary, time-like vectors. As
shown in \cite{Gero2003}, the test particle then necessarily moves
along a geodesic. While this proof do not explicitly include the
cosmological constant term, replacing $\tilde{G}_{\mu\nu} = G_{\mu\nu}
+\Lambda_{DE}g_{\mu\nu}$ does not materially change the nature of the
proof given in \cite{Gero2003}; since $\Lambda_{DE}>0$,
$\tilde{G}{\mu\nu}$ satisfies the strong energy condition as long as
$G_{\mu\nu}$ does. 

If $T_{\mu\nu}=0$ then $\mathfrak{R}[4+8\pi
T/\Lambda_{DE}]=\mathfrak{R}[4]$, and is a constant.  We then see
explicitly from Eq.~$(\ref{genEOM})$ that the 
extended GEOM reduces to the GEOM. As the $T_{\mu\nu}=0$ case is
precisely the situation covered by \cite{Gero2003}, the extended GEOM
does not violate these theorems. Indeed, in the following we will
explicitly construct the energy-momentum tensor for dust within the
extended GEOM framework. 

Consider a collection of massive particles that can be treated as an 
inviscid fluid with density $\rho$, pressure $p$, and fluid velocity,
$v^\mu(x)$. We consider the spacetime to be spatially symmetric, so
that the most general form that the energy-momentum
tensor for this fluid is the usual
\begin{equation}
T_{\mu\nu} = \rho v_\mu v_\nu -\left(g_{\mu\nu}-\frac{v_\mu
  v_\nu}{c^2}\right) p. 
\label{EM}
\end{equation}
We emphasize that this form for $T_{\mu\nu}$ depends only on the
spatial isotropy of the fluid, and thus holds for both the GEOM
and the extended GEOM. 

Following \cite{MTW}, energy and momentum conservation,
$\nabla^\nu T_{\mu\nu}=0$, requires that  
\begin{equation}
0 = v_\nu \nabla^\nu(\rho+p/c^2)v_\mu + (\rho+p/c^2)\nabla_\nu v^\nu v_\mu +
(\rho+p/c^2) v^\nu\nabla_\nu v_\mu - \nabla_\mu p.
\label{Euler}
\end{equation}
Since $v^2 = $ constant even within the extended GEOM
formulation, projecting the above along $v_\mu$ gives once again the
first law of thermodynamics
\begin{equation}
d(V\rho c^2) = - pdV,
\label{thermo}
\end{equation}
where $V$ is the volume of the fluid. This analysis holds for both the
GEOM and the extended GEOM, and thus the first law of thermodynamics
holds for both equations of motion. The standard analysis of the
evolution of the universe under the extended GEOM therefore follows
much in the same way as before.  

Next, projecting Eq.~$(\ref{Euler})$ along the subspace
perpendicular to $v_\mu$ gives the relativistic version
of Euler's equation 
\begin{equation}
0 = \left(\rho+\frac{p}{c^2}\right) v^\nu \nabla_\nu v_\mu - \left(g_{\mu\nu}-\frac{v_\mu
v_\nu}{c^2}\right)\nabla^\nu p.
\label{spatial}
\end{equation}
Once again, Eq.~$(\ref{spatial})$ holds for both sets of equations of
motion. 

Consider now the simplest case when the constituent test
particles in the fluid do not interact with one another except under
gravity. This corresponds to the case of ``dust''. If test particles
in this dust follow the GEOM, the solution to Eq.~$(\ref{spatial})$
gives the usual $T_{\mu\nu}^{\hbox{\scriptsize{GEOM-Dust}}} = \rho
v_\mu v_\nu$ with $p\equiv 0$. If, on the other hand, the test
particles follow the extended GEOM, the situation changes. Using 
Eq.~$(\ref{genEOM})$, Eq.~$(\ref{spatial})$ becomes 
\begin{equation}
0=\left(g_{\mu\nu}-\frac{v_\mu v_\nu}{c^2}\right)\left\{(\rho
c^2+p)\nabla^\nu \log\mathfrak{R} - \nabla^\nu p\right\}, 
\end{equation}
so that 
\begin{equation}
(\rho c^2+p)\nabla_\mu \mathfrak{R} - \mathfrak{R}\nabla_\mu p=\xi_{DE}
  \>\Lambda_{DE}\> cv_\mu,
\label{p}
\end{equation}
where $\xi_{DE}$ is a constant. By contracting the above with $v_\mu$, it
is straightforward to see that if $\xi_{DE}\ne0$, $p$ will increase
linearly with the proper time. This would be unphysical, and we
conclude that $\xi_{DE}$ must be zero. 

Taking the pressure as a function of only the density,
Eq.~$(\ref{p})$ reduces to a nonlinear, first order, ordinary
differential equation. We will not solve this equation in general.
Instead, we look at the nonrelativistic limit where $\rho c^2 \gg
3p$. Then $T\approx \rho c^2$, and $\mathfrak{R}$ can be approximated
as a function of $\rho$ only. Equation $(\ref{p})$ can then be 
solved implicitly to give
\begin{equation}
p(\rho) = -\rho c^2 + c^2\mathfrak{R}[4+8\pi\rho/\Lambda_{DE}]\int_0^\rho
\frac{ds}{\mathfrak{R}[4+8\pi s/\Lambda_{DE}]}.
\label{pressure}
\end{equation}
Given the density, the pressure is then determined once a form for
$\mathfrak{R}$ is known. The energy-momentum tensor for dust,
$T^{\hbox{\scriptsize{Ext-Dust}}}$, under the extended GEOM can then
be constructed using $\rho$, and the resultant $p$ from
Eq.~$(\ref{pressure})$. Note also from Eq.~$(\ref{pressure})$ that
$p\approx 4\pi\mathfrak{R}'[4]\rho^2c^2/\mathfrak{R}[4]\Lambda_{DE}$ when
$\rho \to 0$, while $p\approx -
\Lambda_{DE}c^2\left(1-1/\mathfrak{R}[4]\right)$ when $\rho\gg
\Lambda_{DE}/2\pi$. Thus, $T^{\hbox{\scriptsize{Ext-dust}}}_{\mu\nu}
\approx \rho v_\mu v_\nu$, and the solution Eq.~$(\ref{pressure})$ is
consistent with the approximation $T\approx \rho c^2$.

The physical reason for the presence of this pressure term in
$T^{\hbox{\scriptsize{Ext-Dust}}}_{\mu\nu}$ can be
seen from Eqs.~$(\ref{genEOM})$ and $(\ref{extendMass})$.  If the second 
equation is solved for the energy of a particle, it is straightforward
to see that under the extended GEOM a collection of test particles behave
as though they were in an external potential set by
$\mathfrak{D}(c^2R/\Lambda_{DE}G)$.  As such, the particles no longer
follow geodesics as they do in the GEOM even though they only interact
with each other through gravity.  The presence of the nonzero pressure
term in $T^{\hbox{\scriptsize{Ext-Dust}}}_{\mu\nu}$ is a reflection of
the presence of this potential. 

\section{A Form for $\mathfrak{D}(x)$ and Experimental Bounds on
  $\alpha_\Lambda$}   
\label{sec:9}

Our analysis up to now is valid for all $\mathfrak{D}(x)$. Requiring that
our extension of the GEOM does not produce effects that would
have already been observed in experiments will fix a
specific form for $\mathfrak{D}(x)$. 

Since our extension of the GEOM does not change the equations of
motion for massless test particles, we expect Eq.~$(\ref{genEOM})$ to
reduce to the GEOM in the ultrarelativistic limit. It is thus only in
the \textit{nonrelativistic} limit where the effects of the deviations
from the GEOM due to the additional terms in Eq.~$(\ref{genEOM})$ can
be seen. We therefore focus on the impact of the extension in the
nonrelativistic, weak gravity limit, and begin by expressing
Eq.~$(\ref{genEOM})$ in these limits.   

\subsection{Constructing $\mathfrak{D}(x)$}
\label{sec:10}

We first perturb off the Minkowski metric $\eta_{\mu\nu}$ in the weak
gravity limit by taking $g_{\mu\nu} =  \eta_{\mu\nu} + h_{\mu\nu}$,
where the only nonzero component of $h_{\mu\nu}$ is
$h_{00}=2\Phi/c^2$, and $\Phi$ is the gravitational
potential. Equation $(\ref{EinsteinEquation})$ then gives 
\begin{equation}
\mathbf{\nabla}^2\Phi + 2\frac{\Lambda_{DE}G}{c^2}\Phi = 4\pi\rho G -
\Lambda_{DE}G, 
\label{PhiEOM}
\end{equation}
in the presence of a cosmological constant. Next, the temporal
coordinate, $x^0$, for the extended GEOM in this limit will, as usual,
be approximated by $ct$ to lowest order in $\vert \mathbf{v}\vert/c$. The
spatial coordinates, $\mathbf{x}$, on the other hand, reduce to   
\begin{equation}
\frac{d^2 \mathbf{x}}{dt^2} = -\mathbf{\nabla} \Phi - \left(\frac{4\pi
     c^2}{\Lambda_{DE}}\right)
     \left[\frac{\mathfrak{D}'(4+8\pi\rho/\Lambda_{DE})}{1+
     \mathfrak{D}\left(4 + 8\pi\rho/\Lambda_{DE}\right)}\right]
     \mathbf{\nabla}\rho.
\label{NREOM}
\end{equation}
Here, we have assumed that the spacetime is spatially symmetric, and
that the particle moves through an ambient, nonrelativistic fluid with
density $\rho$.  

For the additional terms in Eq.~$(\ref{NREOM})$ from the extension
\textit{not} to contribute significantly to Newtonian gravity under
current  experimental conditions, $\mathfrak{D}'(4 + 
8\pi\rho/\Lambda_{DE}) \to 0$ when $\rho >> \Lambda_{DE}/2\pi$. We can
now either choose $\mathfrak{D}'(x)>0$ or $\mathfrak{D}'(x)<0$, as
$\mathfrak{D}'(x)\to0$ from either above or below. This is an
ambiguity in the construction of $\mathfrak{D}(x)$ that cannot be
resolved with our current analysis; for now, we take
$\mathfrak{D}'(x)<0$. The simplest form for $\mathfrak{D}'(x)$
with the correct asymptotic properties is    
\begin{equation}
\mathfrak{D}'(x) = -\frac{\chi}{1 + x^{1+{\alpha_\Lambda}}},
\label{D}
\end{equation}
where $\alpha_\Lambda$ is a power-law exponent that determines how
fast $\mathfrak{D}'(x)\to0$ for $x\to\infty$, and $\chi$ is the
normalization constant
\begin{equation}
\frac{1}{\chi} = \int_0^\infty \frac{ds}{1+s^{1+{\alpha_\Lambda}}}.
\end{equation}
To ensure nonzero effective masses, $\mathfrak{D}(x)$ must be
positive, and we integrate Eq.~$(\ref{D})$ to get
\begin{equation}
\mathfrak{D}(x) = \chi(\alpha_\Lambda)\int_x^\infty
\frac{ds}{1+s^{1+{\alpha_\Lambda}}},
\end{equation}
with the condition that ${\alpha_\Lambda} >0$ to ensure that the the
integral is defined. With this choice of integration constants,
$\mathfrak{D}(0)=1$ and $\mathfrak{D}(x) \to 0$ as $x\to
\infty$. While the precise form of $\mathfrak{D}(x)$ is 
calculable, we will not need it. Instead, because
$8\pi\rho/\Lambda_{DE}\ge 0$,   
\begin{equation}
\mathfrak{D}(4+8\pi\rho/\Lambda_{DE}) = \chi\sum_{n=0}^\infty
  \frac{(-1)^n}{n(1+{\alpha_\Lambda})+{\alpha_\Lambda}}
  \left(4+\frac{8\pi\rho}{\Lambda_{DE}}\right)^{-n(1+{\alpha_\Lambda})-{\alpha_\Lambda}},
\label{D-expand}
\end{equation}
while
\begin{equation}
\frac{1}{\chi}=1+2\sum_{n=0}^\infty
\frac{(-1)^n}{[1+(n+1)(1+{\alpha_\Lambda})][n(1+{\alpha_\Lambda})+{\alpha_\Lambda}]}.
\end{equation}
Notice that in the ${\alpha_\Lambda}\to\infty$ limit, $\mathfrak{D}(x)\to
0$, $\mathcal{L}_{\hbox{\scriptsize{Ext}}}\to\mathcal{L}_0$, and the
GEOM is recovered. 

Bounds on $\alpha_\Lambda$ will be found below. For now, we note that
for ${\alpha_\Lambda} > 1$, $\chi\sim 1$ and 
$\mathfrak{D}(4+8\pi\rho/\Lambda_{DE}) \approx 0$. Thus, 
\begin{equation}
\frac{d^2 \mathbf{x}}{dt^2} \approx -\mathbf{\nabla} \Phi + \left(\frac{4\pi
     c^2\chi}{\Lambda_{DE}}\right)
     \left\{1+\left(4+\frac{8\pi\rho}{\Lambda_{DE}}\right)^{1+{\alpha_\Lambda}}\right\}^{-1}  
     \mathbf{\nabla}\rho. 
\label{FinalEOM}
\end{equation}

\subsection{Constraints From Terrestrial Experiments}
\label{sec:11}

We now consider the constraints placed on $\mathfrak{D}(x)$ due to
terrestrial experiments. 

From WMAP, $\Lambda_{DE}= 7.21_{-0.84}^{0.82} \times 10^{-30}$
g/cm${}^3$, which for hydrogen atoms corresponds to a number density
of $\sim 4$ atoms/m${}^3$. It is clear that the density of both solids and
liquids far exceed $\Lambda_{DE}$, and in such media
Eq.~$(\ref{FinalEOM})$ reduces to what one expects for Newtonian
gravity. Only very rare gases, in correspondingly hard vacuums,
can have a density that is small enough for the additional terms in
Eq.~$(\ref{NREOM})$ to be relevant. The hardest vacuum currently
attainable experimentally is $\sim 10^{-13}$ torr
\cite{Ishi}. For a gas of He${}_4$ atoms at 4 $^{o}$K,
this corresponds to a density of $\rho_{\hbox{\scriptsize limit}}
\approx 10^{-18}$ g/cm${}^3$, which is 11 orders of magnitude smaller
than $\Lambda_{DE}$. Nevertheless, because the scale of the
acceleration from the additional terms in
Eq.~$(\ref{FinalEOM})$\textemdash which goes as $c^2/L$ for the
relevant length scale $L$ in the experiment\textemdash is so
large, effects at these densities will nevertheless be
relevant. Indeed, we will use $\rho_{\hbox{\scriptsize limit}}$ to
determine a lower bound to $\alpha_\Lambda$. 

Consider a simple experiment that looks for signatures of the
extension of the GEOM Eq.~$(\ref{FinalEOM})$ by looking for 
anomalous accelerations (through pressure fluctuations) in a gas of
He${}^4$ atoms at 4 $^o$K with a density
$\rho_{\hbox{\scriptsize{limit}}}$. Inside this gas we consider a
sound wave with amplitude $\epsilon \rho_{\hbox{\scriptsize limit}}$
propagating with a wavenumber $k$. Suppose that the smallest
measurable acceleration for a test particle in this gas is
$a_{\hbox{\scriptsize bound}}$. For the additional terms in
Eq.~$(\ref{FinalEOM})$ to be undetectable,  
\begin{equation}
a_{\hbox{\scriptsize{bound}}} \ge \frac{c^2\chi}{2}
\left(\frac{\Lambda_{DE}}{8\pi\rho_{\hbox{\scriptsize{limit}}}}\right)^{\alpha_\Lambda} \epsilon k.
\end{equation}
This gives a lower bound on ${\alpha_\Lambda}$ as
\begin{equation}
{\alpha_\Lambda}_{\hbox{\scriptsize{bound}}} = \frac{\log{\left[
      2a_{\hbox{\scriptsize{bound}}}/c^2 \chi\epsilon k
      \right]}}{
    \log{\left[\Lambda_{DE}/8\pi\rho_{\hbox{\scriptsize{limit}}}\right]}}.
\label{alphaBound}
\end{equation}
For $\epsilon =0.1$, $k=1$ cm${}^{-1}$, and
$a_{\hbox{\scriptsize{bound}}} = 1$ cm/s${}^2$, 
${\alpha_\Lambda}_{\hbox{\scriptsize{bound}}}$ ranges from $1.28$ for
$\Lambda_{DE} = 10^{-32}$ g/cm${}^3$ to $1.58$ for $\Lambda_{DE} =
10^{-29}$ g/cm${}^3$. Notice that because 
${\alpha_\Lambda}_{\hbox{\scriptsize{bound}}}$ depends only
logarithmically on $\epsilon$, $k$, and $a_{\hbox{\scriptsize{bound}}}$,
the lower bound for $\alpha_\Lambda$ is relatively insensitive to
the specific values taken for these parameters. 

Adelberger has recently done state-of-the
art, E\"ovtos-type experiments \cite{Adel2001}, \cite{Adel2003},
\cite{Adel2004} to test the Newtonian, $1/r^2$ law for gravity. While
the pressures under which these experiments were performed were not
explicitly stated, as far as we know these experiments were not done
at pressures lower than $10^{-13}$ torr; we thus would not expect effects
from extended GEOM to be apparent in these experiments either.

\subsection{Constraints from Solar System Observations}
\label{sec:12}

Most of the classical tests of general relativity are based on the
motion of bodies in the solar system. We now consider the constraints
placed on $\mathfrak{D}(x)$ due to these tests.

We begin by noting that in idealized situations such as the analysis
of the advancement of perihelion of Mercury, the energy-momentum tensor is 
taken to be zero outside of a massive body such as the Sun; the terms
on the right hand side of Eq.~$(\ref{genEOM})$ will clearly not affect these
analyses. While this argument would seem to hold for this and the
other experimental tests of general relativity, it is much too
simplistic. In practice, the $T_{\mu\nu}$ in each of these tests does
not, in fact, vanish; a background density is always present. Except
for experiments involving electromagnetic  
waves, what is instead needed is a comparison of the background
density with $\Lambda_{DE}$. It is only when this density is much
greater than $\Lambda_{DE}/2\pi$ that the additional terms in
Eq.~$(\ref{genEOM})$ may be negligible. We present this analysis below.    

Of the classical, solar system tests of general relativity, only in
observations or experiments involving motion of massive
test particles can effects of the extension be seen. The extended 
GEOM does not affect the motion of electromagnetic waves, and such tests of
general relativity as the gravitational redshift and the deflection of
light by the Sun will not be affected by the extension. Considering
first the advancement of the perihelion of Mercury, we note that the
density of matter, $\rho_{\hbox{\scriptsize{orb}}}$, (due primarily
to the solar wind) in the region of Mercury's orbit is $\sim 10^{-23}$
g/cm$^3$. This density is only five orders of magnitude smaller to
$\rho_{\hbox{\scriptsize{limit}}}$. There is, then, a possibility that
the extended GEOM could change the amount in which the perihelion of
Mercury advances due to general relativistic effects. 

To show that our extension nevertheless does not change the advance of
perihelion, we follow extensively the derivation of the
advancement of perihelion given in \cite{MTW}. In particular, we use
a Hamilton-Jacobi-based analysis to calculate the advancement and
replace  
Eq.~$(\ref{extendMass})$ by  
\begin{eqnarray}
m^2_{\hbox{\scriptsize{M}}}c^2\left[1+
  \mathfrak{D}\left(4+\frac{8\pi\rho_{\hbox{\scriptsize{orb}}}}{\Lambda_{DE}}\right)\right]
&=&\left[1+2\frac{GM_{\bigodot}}{c^2r}+
  2\left(\frac{GM_{\bigodot}}{c^2r}\right)^2\right] 
\left(\frac{\partial S_{\hbox{\scriptsize{M}}}}
     {c\partial t}\right)^2 
-
\nonumber
\\
&\>&\left[1-2\frac{GM_{\bigodot}}{c^2r}\right]
\left\{\left(\frac{\partial S_{\hbox{\scriptsize{M}}}}{\partial
  r}\right)^2 
+\left(\frac{1}{r}\frac{\partial
  S_{\hbox{\scriptsize{M}}}}{\partial \phi}\right)^2\right\}.
\nonumber
\\ 
\label{HJ}
\end{eqnarray}
Here, $m_{\hbox{\scriptsize{M}}}$ is the mass of Mercury,
$M_{\bigodot}$ is the mass of the Sun, and
$S_{\hbox{\scriptsize{M}}}$ is the action for the motion of Mercury
about the Sun. We have used the usual Schwarzchild solution for the
region outside of the Sun, and  kept terms up to second order in
$GM/c^2r$.  While in principal we would expect the extended GEOM to
change the solution of Einstein's equations, we are working in both
the weak gravity and nonrelativistic limits, where, as we have shown
above, $T^{\hbox{\scriptsize{Ext-Dust}}}\approx
\rho_{\hbox{\scriptsize{orb}}}c^2$; the Schwarzchild solution
is unchanged in these limits. We also note that because
$\rho_{\hbox{\scriptsize{orb}}}\sim 10^{-23}$ g/cm$^{3}$, the
additional matter in Mercury's orbit will not affect the form of the
Schwarzchild solution, and can be neglected when determining the metric
outside of the Sun. 

We next look for solutions of the following form:  
\begin{equation}
S_{\hbox{\scriptsize{M}}} = -m_{\hbox{\scriptsize{M}}}c^2\tilde{E}t+
m_{\hbox{\scriptsize{M}}}\tilde{L}\varphi +
m_{\hbox{\scriptsize{M}}}c\>\mathfrak{S}(r),
\label{action}
\end{equation}
where $\tilde{E}$ is the energy per unit rest energy,
$\tilde{L}$ is the angular momentum per unit rest mass, $\varphi$ is
the azimuthal angle, and $r$ is the radial position of Mercury
measured from the Sun. Solution of Eq.$(\ref{HJ})$ then gives
\begin{eqnarray}
\mathfrak{S} &=& \pm \int^r\Bigg\{-(1+\mathfrak{D}-\tilde{E}^2) +
2\frac{GM}{c^2r}\left[1-2(1-\tilde{E}^2)\right]-
\nonumber
\\
&{}& 
\qquad
\frac{\tilde{L}^2}{c^2r^2}+6\left(\frac{GM}{c^2r}\right)^2\Bigg\}^{1/2}dr,
\label{phase1}
\end{eqnarray}
after keeping terms up to order $(GM_{\bigodot}/c^2r)^2$. In obtaining
Eq.~$(\ref{phase1})$, we have used the fact that $1 - \tilde{E}^2 \sim
GM_{\bigodot}/c^2a$, where $a$ is the semi-major axis of Mercury's
orbit; observations set $a=57.91\times 10^6$ km \cite{NASA}, giving
$1-\tilde{E}^2 \sim 10^{-8}$. We next used the relation from
Eq.~$(\ref{D-expand})$,
$\mathfrak{D}[4+8\pi\rho_{\hbox{\scriptsize{orb}}}/\Lambda_{DE}] 
  \approx
  (\chi/\alpha_\Lambda)
  \left(\Lambda_{DE}/8\pi\rho_{\hbox{\scriptsize{orb}}}\right)^{\alpha_\Lambda}$. 
Thus, $\mathfrak{D} \sim 10^{-10} - 10^{-12}$ for $\alpha_\Lambda = 1.28 - 
  1.58$, so that $GM_{\bigodot}/c^2a < \mathfrak{D} <
  (GM_{\bigodot}/c^2a)^2$. 

The shape of the orbit of Mercury is determined by minimizing the action,
$\partial S/\partial \tilde{L} = 0$, in Eq.~$(\ref{action})$. The
resultant integral is straightforward to calculate, and we obtain
\begin{equation}
\varphi =
\left[1+3\left(\frac{GM_{\bigodot}}
     {c\tilde{L}}\right)^2\right]\cos^{-1}\left[\frac{(1-e^2)a}{er}-\frac{1}{e}\right],
\label{phi}
\end{equation}
where analytically
\begin{eqnarray}
a &=&
\frac{GM_{\bigodot}}{c^2}\frac{1-2(1-\tilde{E}^2)}{1+\mathfrak{D}-\tilde{E}^2},
\nonumber
\\
1-e^2 &=& \left(\frac{c\tilde{L}}{GM_{\bigodot}}\right)^2
  \left(1+\mathfrak{D}-\tilde{E}^2\right)
  \left[1+4(1-\tilde{E}^2)-6\left(\frac{GM_{\bigodot}}{c\tilde{L}}\right)^2\right],
\label{ae}
\end{eqnarray}
and $e$ is the eccentricity of the orbit. The advancement in the
perihelion of Mercury therefore still has the form 
\begin{equation}
\delta\varphi = \frac{6\pi GM_{\bigodot}}{c^2a(1-e^2)}.
\end{equation}
Note, however, that the product $a(1-e^2)$ is independent
of $\mathfrak{D}$. \textit{Thus, the amount that the perihelion of Mercury's
orbit advances due to general relativistic effects is not changed by
the extended GEOM.} Physically, this is because the extended GEOM
only modifies the rest mass in Eq.~$(\ref{extendMass})$, and does not
modify terms explicitly dependent on the velocity of Mercury.

What the extended GEOM does affect is the analytical expressions for
$e$ and $a$. However, in calculating the numerical value of
$\delta\varphi$, both $e$ and $a$ are taken as \textit{measured}
quantities obtained from observations; they are not 
calculated from first principles. The fact that there is now a slight
different relationship between $e$ and $a$, and the total energy,
$m_{\hbox{\scriptsize{M}}}c^2\tilde{E}$, and the angular momentum,
$m_{\hbox{\scriptsize{M}}}\tilde{L}$, of Mercury (by less than 0.01
\% for $\alpha_\Lambda = 1.58$) would require an independent method of
determining $\tilde{E}$ and $\tilde{L}$ to check. Such an independent
determination is not currently available.  

There are three other potential solar system tests of general
relativity that could, in principal, be used to constrain the extended
GEOM. The first is looking at three-body effects in the lunar 
orbit due to the Sun, Earth and the Moon. It is questionable whether
such effects can be measured \cite{MTW}, however, and they therefore
cannot serve as a check on the extended GEOM. The second is frame
dragging, where the precession of a spinning 
object is caused azimuthal changes in the local metric of spacetime by
the rotating Earth; this effect is being measured by Gravity Probe B
\cite{grav-B}. The third also measures frame dragging, but uses two
satellites\textemdash LAGEOS and LAGEOS 2\textemdash instead of a
spinning body, to detect the Lense-Thirring effect; frame dragging was
recently seen \cite{FrameDrag} in this experiment.  

Frame dragging, whether it is by a spinning object or by two orbiting
satellites, is inherently a velocity-dependent effect, however, that
couples either to the spin of the object or to the orbital angular
momentum of the satellites with the spacetime metric. Our extension of
the GEOM changes the rest mass of the test particle, and in a spherically
symmetric geometry in the nonrelativistic limit, the additional effects
due to the extension is radial. We thus would not expect the effects
from our extension can be seen either from the precession of a spin,
or through the Lense-Thirring effect.  

\section{Concluding Remarks}
\label{sec:13}

We have shown that because of the existence of a universal length scale,
$\ell_{DE}$, it is now possible to construct an extension of the
GEOM. This extension preserves the equivalence principal, does not
change the motion of massless test particles, and does not produce effects
that would be detectable in either terrestrial experiments, or through
observing the motion of bodies in the solar system. Our extension of
the GEOM is thus a physically viable alternative to the GEOM. 

The question remains as to whether these equations of motion have any
physical relevance. In short, is anything gained by using this
extension? Because $\ell_{DE}=14010^{800}_{820}$ Mpc, we would expect
that any effects from the extended GEOM will become apparent at much
longer length scales than those considered here. Indeed, given the size of
$\ell_{DE}$ the only reason why we would expect the extended GEOM to
be relevant at all is because $\mathfrak{D}$ is a nonlinear function
of the energy-momentum tensor of ambient matter. This question of
relevance will be addressed in a future paper \cite{ADS} where the
extended GEOM is applied to the motion of stars in the rotation curves
of galaxies, and to the density of matter at cosmological length
scales. These are the scales at which we expect the effects from
the extension to come into play, and where its relevance can be assessed.   

Finally, as noted in \cite{MTW}, Eq.~$(\ref{spatial})$ can be
solved in general to give an equation of motion for particles, $v^\nu
\nabla_\nu v_\mu = \nabla_\mu p/(\rho+p/c^2)$, and we see that the
presence of any pressure term in the energy-momentum tensor results in
deviations from geodesic motion.  Given that $\Lambda_{DE}$ can also
be used to construct a pressure, it is natural to ask whether the
effects of the extended GEOM can be obtained through the introduction
of an ad hoc pressure term in the energy-momentum tensor. Such an ad hoc term
can only be introduced to the energy-momentum tensor for matter,
however; for the reasons given in the introduction, the equations of
motion for massless particles cannot be changed. In addition, because the
behavior of any massive particle approaches that of a massless one in
the ultrarelativistic limit, this ad hoc pressure term must be
constructed so that irrespective of frame this term contributes a
negligible amount to the energy of the particle in this
limit. Moreover, even if such a construction can be accomplished at the
ultrarelativistic limit, hurdles remain at the nonrelativistic limit.
While it is possible to construct in the nonrelativistic limit an
appropriate ad hoc pressure using $\Lambda_{DE}$ and $\mathfrak{D}$,
when the resultant equations of motion are applied to the same systems as
the extended GEOM in \cite{ADS}, effects are predicted at the
cosmological scale that either do not agree with experiment or are
not physically reasonable. This occurs even though the pressure is
chosen so that at galactic scales predicted effects will be in broad
agreement with observations. For these reasons, it is doubtful that
introducing ad hoc pressure term in place of the extended GEOM
will be succussful.

\begin{acknowledgements}

The author would like to thank John Garrison for the numerous suggestions,
comments, and the support he has given of his time while this research
was being done. His efforts have helped guide it, and have elucidated
many of the arguments given here. The author would also like to thank
K.-W. Ng, H. T. Cho, and Clifford Richardson for their comments and
criticisms while this research was done.

\end{acknowledgements}

\end{document}